%% file: main.tex
\documentclass[sigconf]{acmart}

\usepackage{tabularray}
\UseTblrLibrary{booktabs}
\usepackage{multirow}
\usepackage{graphicx}
\usepackage{balance}  % for  \balance command ON LAST PAGE  (only there!)
\usepackage{booktabs} % For formal tables
\usepackage{url}
\usepackage{xcolor}
\usepackage{subcaption}
\usepackage[ruled,vlined,linesnumbered]{algorithm2e}
\usepackage{listings}
\usepackage{fancyvrb}
\usepackage{mathtools}
\usepackage{amsmath}
\usepackage{enumitem}
\usepackage{float}

\AtBeginDocument{%
  }

\setcopyright{acmlicensed}
\copyrightyear{2026}
\acmYear{2026}
\acmDOI{XXXXXXX.XXXXXXX}
\acmConference[CIDR'26]{The Conference on Innovative Data Systems Research}{January 18--21,
  2026}{Chaminade, USA}

\acmISBN{978-1-4503-XXXX-X/2018/06}
\newcommand{\spara}[1]{\vspace*{0.05in} \noindent \textbf{#1.}}

\begin{document}

\title{Rethinking Analytical Processing in the GPU Era}
%\LARGE \textit{Building Sirius: A GPU-Native SQL Engine}}
% Your text here

\author{
Bobbi Yogatama$^2$$^*$, 
Yifei Yang$^1$$^*$, 
Kevin Kristensen$^1$, 
Devesh Sarda$^1$, 
Abigale Kim$^1$, 
Adrian Cockcroft$^5$, 
Yu Teng, 
Joshua Patterson$^2$, 
Gregory Kimball$^2$, 
Wes McKinney$^4$, 
Weiwei Gong$^3$, 
Xiangyao Yu$^1$
}

% \author{
% Bobbi Yogatama\textsuperscript{1}\textsuperscript{*}, 
% Yifei Yang\textsuperscript{1}\textsuperscript{*}, 
% Kevin Kristensen\textsuperscript{1}, 
% Devesh Sarda\textsuperscript{1}, 
% Abigale Kim\textsuperscript{1}, 
% Adrian Cockcroft\textsuperscript{5}, 
% Yu Teng, 
% Joshua Patterson\textsuperscript{2}, 
% Gregory Kimball\textsuperscript{2}, 
% Wes McKinney\textsuperscript{4}, 
% Weiwei Gong\textsuperscript{3}, 
% Xiangyao Yu\textsuperscript{1}
% }

% \authornote{Equal contributions}

\affiliation{
\textsuperscript{1}University of Wisconsin-Madison, 
\textsuperscript{2}NVIDIA, 
\textsuperscript{3}Oracle, 
\textsuperscript{4}Posit PBC,
\textsuperscript{5}OrionX
}

\email{siriusdb@cs.wisc.edu}

\begin{abstract}
The era of GPU-powered data analytics has arrived. In this paper, we argue that recent advances in hardware (e.g., larger GPU memory, faster interconnect and IO, and declining cost) and software (e.g., composable data systems and mature libraries) have removed the key barriers that have limited the wider adoption of GPU data analytics. 
We present Sirius, a prototype open-source GPU-native SQL engine that offers drop-in acceleration for diverse data systems. 
Sirius treats GPU as the primary engine and leverages libraries like libcudf for high-performance relational operators. It provides drop-in acceleration for existing databases by leveraging the standard Substrait query representation, replacing the CPU engine without changing the user-facing interface. 
% On TPC-H, Sirius achieves 8.3$\times$ speedup when integrated with DuckDB in a single node at the same hardware rental cost, and up to 12.5$\times$ speedup when integrated with Apache Doris in a distributed setting. 
% On Clickbench, Sirius achieves 7.2$\times$ lower total cost of ownership compared to the top submission in Clickbench.
Sirius achieves 8.3$\times$ and 7.4$\times$ better cost efficiency on TPC-H and ClickBench, respectively, when integrated with single-node DuckDB, and delivers up to 12.5$\times$ speedup when integrated with Apache Doris distributed engine.

\renewcommand{\thefootnote}{\fnsymbol{footnote}}
\footnotetext[0]{$^*$Equal contributions}
\renewcommand{\thefootnote}{\arabic{footnote}}
\end{abstract}

\settopmatter{printacmref=false}

\maketitle

\input{introduction}
\input{motivation}
\input{sirius}

\input{evaluation}
\input{conclusion}
\input{acknowledgment}

\bibliographystyle{ACM-Reference-Format}
\bibliography{bib/sirius.bib}
\end{document}

%% file: introduction.tex
% \vspace{-.1in}
\section{Introduction}
\label{sec:introduction}

The performance of a SQL engine is ultimately driven by the capabilities of the underlying hardware---especially compute throughput and memory bandwidth. Modern GPUs are rapidly outpacing CPUs on both fronts and have become the default hardware choice for data- and compute-intensive applications such as machine learning. Table 1 compares recent CPU and GPU architectures, highlighting the contrast in core count and memory bandwidth. Given the inherently parallel nature of relational data analytics, GPUs are a perfect fit for future analytical databases. 

% \begin{table}[h]
% \centering
% \caption{Comparison of CPU and GPU Instances}
% \vspace{-.1in}
% \begin{tabular}{|l|p{1in}|p{1.1in}|}
% \hline
%   & {\centering Amazon c6a.metal \newline (AMD EPYC CPU)}
%   & {\centering GH200 \newline (NVIDIA GPU)} \\ \hline
% Core Count & 192 (vCPUs) & 14,000+ (CUDA cores) \\
% \hline
% Memory BW & $\sim$400 GB/s & 3,000 GB/s (HBM)\\
% \hline
% Memory Size & 384 GB & 96 GB (HBM) \\
% \hline
% Rental Cost & \$7.344/h (AWS) & \$3.2/h (Lambda Labs) \\
% \hline
% \end{tabular}
% \label{tab:cpu_gpu_comparison}
% \end{table}

% \vspace{.3in}
\begin{table}[t]
  \centering
  \caption{Comparison of CPU and GPU Instances}
  \vspace{-.15in}
  \small
  \begin{tblr}{
    colspec = {c|c|c}
  }
  \toprule
    & \begin{tabular}{@{}c@{}}{Amazon c6a.metal}\\(AMD EPYC CPU)\end{tabular}
    & \begin{tabular}{@{}c@{}}{GH200}\\(NVIDIA GPU)\end{tabular} \\
  \cmidrule{1-3}\morecmidrules\cmidrule{1-3}
  Core Count & 192 (vCPUs) & 14,000+ (CUDA cores) \\
  \cmidrule{1-3}
  Memory BW & $\sim$400 GB/s & 3,000 GB/s (HBM)\\
  \cmidrule{1-3}
  Memory Size & 384 GB & 96 GB (HBM) \\
  \cmidrule{1-3}
  Rental Cost & \$7.344/h (AWS) & \$1.5/h (Lambda Labs) \\
  \bottomrule
  \end{tblr}
  \label{tab:cpu_gpu_comparison}
  \vspace{-.1in}
\end{table}
% \vspace{-.2in}

Despite their great potential, however, GPU-based SQL engines \cite{omnisci, kinetica, theseus, cudf, sparkrapids, theseus} have not yet seen mainstream adoption. Prior efforts, both academic and commercial, have been constrained by the following four key challenges and thereby remain niche solutions. 

\setlist[itemize]{align=parleft,left=0pt..1em}
\begin{itemize}
\item \textbf{Limited GPU memory capacity}: Up to 288~GB in a latest GPU device in contrast to several TBs in CPUs.
\item \textbf{Data movement bottleneck}: Traditionally, PCIe has relatively low bandwidth, creating a bottleneck when transferring data between GPU and the rest of the system. 
\item \textbf{Expensive GPU hardware}: High-end GPUs remain significantly more expensive than CPUs and are often in short supply. 
\item \textbf{Engineering cost}: Building a full SQL engine optimized for GPUs from the ground up is a major engineering challenge.
\end{itemize}

In this paper, we argue that the hardware and software foundations are finally in place today to overcome these challenges and enable scalable GPU data analytics.
On the hardware side, GPU memory capacity doubles almost every generation, starting from Volta (32 GB), to Ampere (80 GB), Hopper (192 GB), and most recently Blackwell (288 GB). Better interconnects such as PCIe Gen6, NVLink-C2C~\cite{nvlinkc2c}, and GPU Direct~\cite{gpudirect} significantly reduce the data movement overhead between GPU and other system components. This allows a GPU to process data beyond on-device memory with very fast speed, enabling TBs of analytics even on a single GPU and more with distribution. 
Meanwhile, GPUs are increasingly affordable and accessible, especially for older generations which are already sufficient for data analytics workloads. 

On the software side, the GPU ecosystem has significantly matured. Libraries such as \texttt{libcudf}~\cite{cudf} provides high-performance primitives like joins and aggregations that a GPU SQL engine can build upon. The rise of composable data systems~\cite{composable}---driven by open standards like Substrait~\cite{substrait} for query representation, Apache Arrow and Parquet for columnar data formats---enable greater interoperability. A modern GPU engine can reuse existing components, such as SQL parser and query optimizer, to drastically reduce the complexity of building the system from the ground up. 

To demonstrate the feasibility of GPU-native databases, we have built Sirius\footnote{https://github.com/sirius-db/sirius}, a prototype open-source SQL engine that uses GPU as the primary execution device and delivers drop-in acceleration across diverse data systems. Sirius integrates seamlessly with existing database systems via the standard Substrait query representation. For example, plugging Sirius into DuckDB causes minimal change to the user-facing interface. Instead of executing the query using DuckDB’s CPU engine, the Substrait plan is routed to Sirius for GPU-native execution. 
Sirius builds on GPU libraries such as \texttt{libcudf}~\cite{cudf}, \texttt{RMM}~\cite{rmm}, and \texttt{NCCL}~\cite{nccl}, reusing optimized implementations of core relational operators like joins, filters, aggregations, and data shuffle. 
Thanks to its modular design, Sirius also allows developers to easily switch the operator implementation between these GPU libraries and custom CUDA kernels.
This flexibility reduces engineering overhead while enabling Sirius to achieve end-to-end GPU acceleration.
% Our detailed evaluation shows that Sirius can deliver 8.3$\times$ speedup as a drop-in accelerator for DuckDB~\cite{duckdb} (single-node) and up to 12.5$\times$ speedup as a drop-in accelerator for Apache Doris~\cite{doris} (distributed) when running TPC-H benchmark.
Our detailed evaluation shows that Sirius delivers 8.3$\times$ and 7.4$\times$ better cost efficiency on TPC-H and ClickBench, respectively, when used as a drop-in accelerator for DuckDB~\cite{duckdb} (single-node), and up to 12.5$\times$ speedup as a drop-in accelerator for Apache Doris~\cite{doris} (distributed).

Sirius is under a permissive Apache-2.0 license and welcomes contributions and collaboration from the database community. We invite researchers and practitioners to build on top of Sirius, with the shared mission of driving the next era of data analytics.

% The rest of the paper is organized as follows:  
% We discuss the recent hardware and software trend in Section~\ref{sec:motivation}.
% Section~\ref{sec:sirius} describes Sirius' design, implementation, and future roadmap.
% Section~\ref{sec:evaluation} evaluates the performance of Sirius and Section~\ref{sec:conclusion} concludes the paper. 
% \yxy{can remove this paragraph if need space}

%% file: motivation.tex
\section{GPU Era for Data Analytics: Why Now}
\label{sec:motivation}

% \begin{figure}[ht!]
%     \includegraphics[width=\columnwidth]{figures/breakdown.pdf}
%     \vspace{-.35in}
%     \caption{Performance Breakdown in Sirius}
%     \vspace{-.15in}
%     \label{fig:breakdown}
% \end{figure}

\begin{figure}[t]
    \centering

    % First row
    \begin{subfigure}[b]{0.48\columnwidth}
        \centering
        \includegraphics[width=\columnwidth]{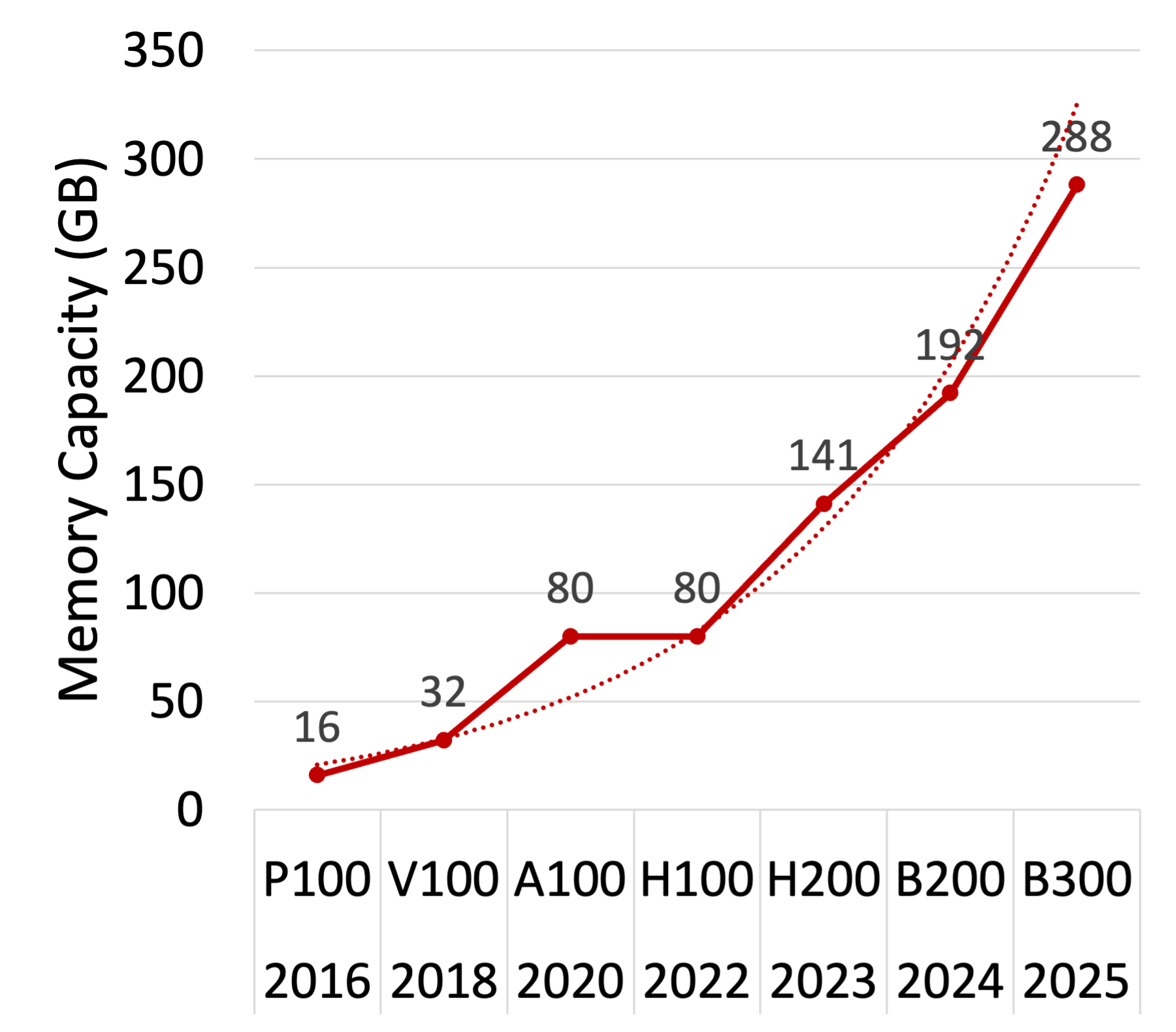}
        \vspace{-.22in}
        \caption{GPU Memory Capacity}
        \vspace{-.1in}
        \label{fig:mem-capacity}
    \end{subfigure}
    \begin{subfigure}[b]{0.48\columnwidth}
        \centering
        \includegraphics[width=\columnwidth]{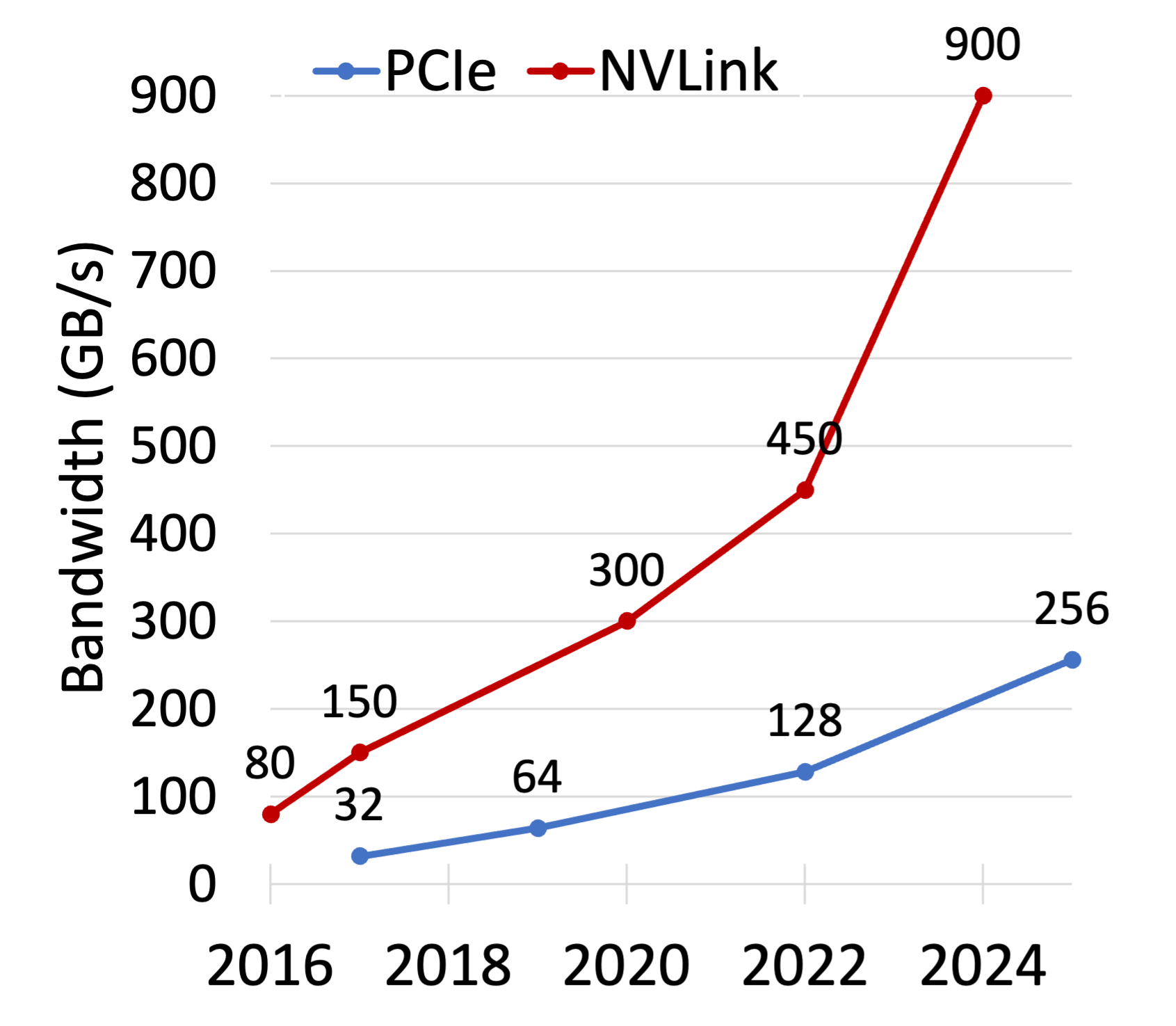}
        \vspace{-.22in}
        \caption{Interconnect Bandwidth}
        \vspace{-.1in}
        \label{fig:interconnect-bw}
    \end{subfigure}
    \begin{subfigure}[b]{0.48\columnwidth}
        \centering
        \includegraphics[width=\columnwidth]{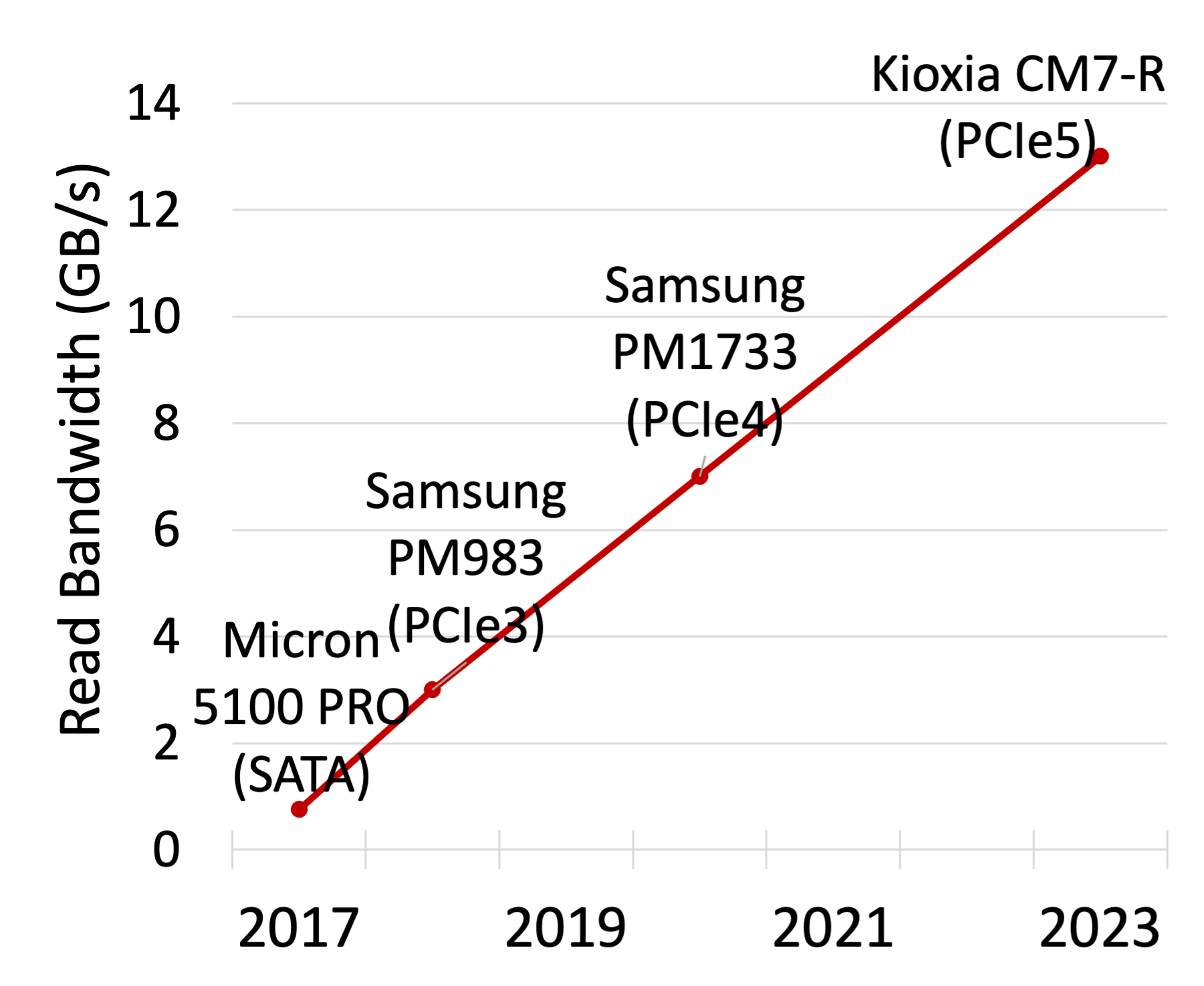}
        \vspace{-.2in}
        \caption{Storage Bandwidth}
        % \vspace{-.1in}
        \label{fig:storage-bw}
    \end{subfigure}
    \begin{subfigure}[b]{0.48\columnwidth}
        \centering
        \includegraphics[width=\columnwidth]{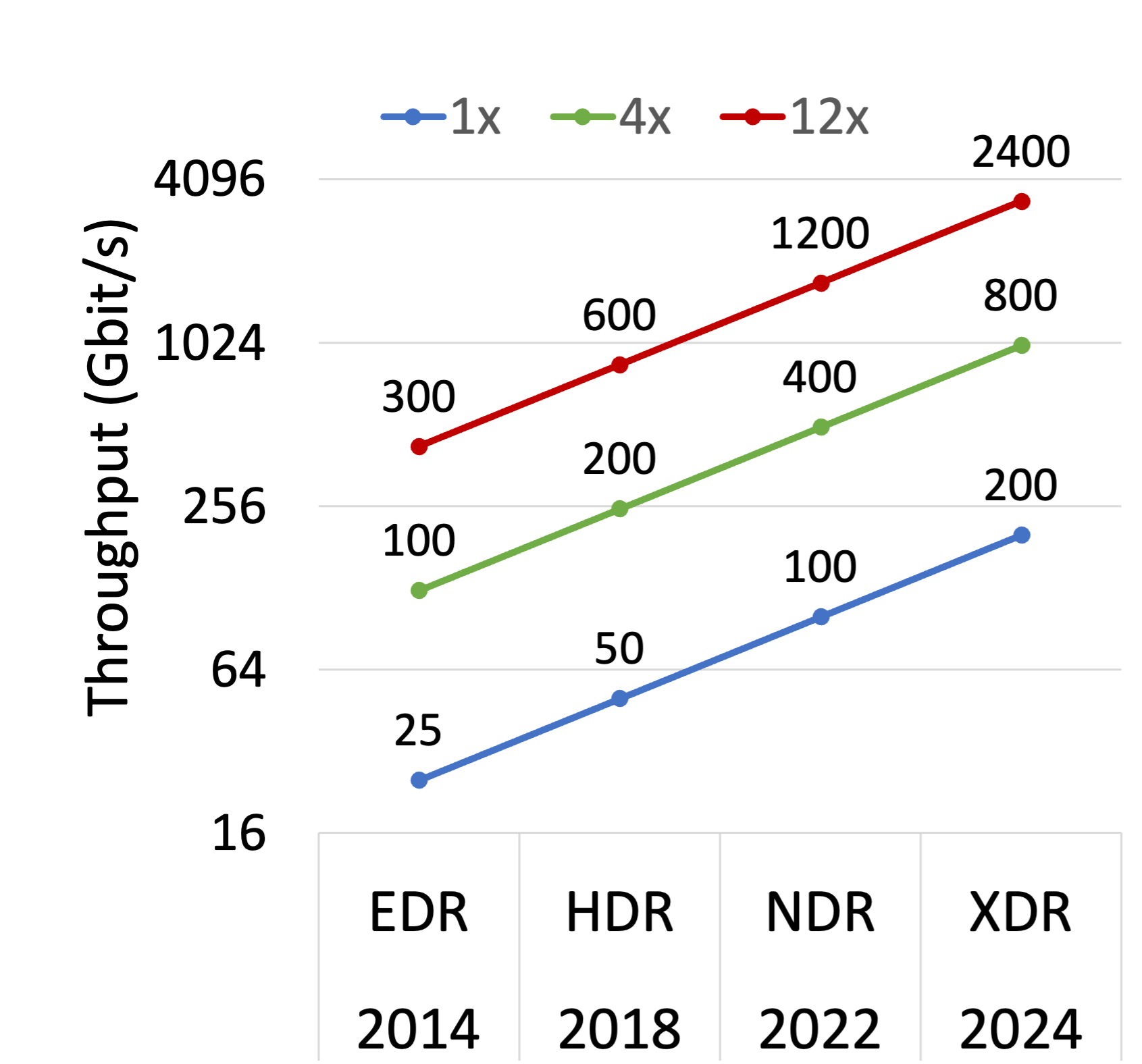}
        \vspace{-.2in}
        \caption{Network Bandwidth}
        % \vspace{-.1in}
        \label{fig:network-bw}
    \end{subfigure}

    \vspace{-.12in}
    \caption{Recent hardware trends}
    \vspace{-.21in}
    \label{fig:hardware-trend}
\end{figure}

This section discusses recent trends in both hardware and software that pave the way towards scalable GPU SQL  analytics. 

\subsection{Hardware Trend}
\label{sec:hardware-trend}

A key barrier to GPU's wider adoption in data analytics has been GPU's limited memory capacity, as well as the limited communication bandwidth between GPU and other system components, such as main memory, storage, and network. In a traditional GPU database, data must fit in GPU memory to fully realize acceleration benefits, which severely limits their use cases. Fortunately, these limitations are diminishing with recent hardware advancements, as illustrated in Figure~\ref{fig:hardware-trend} and discussed below.

\spara{Bigger Memory at Faster Speed} The capacity of GPU device memory has been growing rapidly. While the largest GPU memory was merely 16~GB ten years ago, a modern NVIDIA B300 Ultra or AMD MI350X has 288~GB device memory, which is sufficient for small to medium-sized analytics workloads.

PCIe, the interconnect between GPU and CPU, has traditionally been a severe performance bottleneck due to its relatively low bandwidth. 
However, PCIe bandwidth has been doubling every two years with PCIe Gen6 offering 128 GB/s, which is comparable to CPU memory bandwidth. The NVLink-C2C~\cite{nvlinkc2c} technology takes this even further, enabling 900 GB/s unidirectional bandwidth between CPU and GPU. In a GH200 superchip, the Hopper GPU can access host memory at more than 400 GB/s, which is more than the maximum bandwidth between the CPU and main memory. These faster interconnects allow a GPU analytics engine to support data larger than the GPU device memory, enabling high-speed access to terabytes of main memory and beyond. 

\spara{Faster Network and Storage} 
Besides faster memory speed, modern GPUs also have increasingly better support for network and storage, allowing a GPU engine to go beyond in-memory data processing. Technologies like GPUDirect Storage and GPUDirect RDMA
allow GPUs to access data directly from storage or network with minimal CPU involvement at very high speed. 
Similar to interconnect technology, storage and network bandwidth have also grown rapidly in recent years—and are expected to continue improving (see Figure~\ref{fig:storage-bw} and ~\ref{fig:network-bw}).
Moreover, the recent S3 over RDMA technique further enables high-speed object store for GPUs, which can achieve storage access speed at 200 GB/s~\cite{s3rdma}. 
With significantly higher computational power, GPUs are better positioned than CPUs to fully utilize these high-bandwidth resources.

% The storage and network bandwidth are also growing rapidly (Figure~\ref{fig:storage-bw} and Figure ~\ref{fig:network-bw}), and GPU provides the computation power that can theoretically saturates these bandwidth much better than CPUs.
% Furthermore, modern GPUs also have increasingly better support for network and storage, allowing a GPU engine to go beyond in-memory data processing. Technologies like GPUDirect Storage and GPUDirect RDMA
% allow GPUs to access data directly from storage or network with minimal CPU involvement at very high speed. The recent S3 over RDMA technique further enables high-speed object store for GPUs, which can achieve storage access speed at up to terabytes per second.

% \noindent\textbf{Increased GPU Memory Capacity:} While GPU memory has always been delivering 1--2 orders-of-magnitude higher bandwidth than CPU memory, the GPU memory capacity has been too small for data analytics. [todo]

\spara{Declining GPU Cost} While GPUs of the latest generation remain in high demand with high cost, older generations are becoming increasingly accessible in the cloud at substantially lower cost. For instance, the GH200 instance shown in Table~\ref{tab:cpu_gpu_comparison} has an on-demand price of \$3.2/hour in Lambda Labs, which is cheaper than many high-end CPU instances. 
The on-demand H100 prices in many GPU cloud providers has seen significant price drop from \$8/hour in March 2023 to around \$3/hour in 2025 ~\cite{h100-price-drop1, h100-price-drop2}.
Even in major cloud provider such as AWS, in 2025 alone, the A100, H100, and H200 on-demand price have dropped by 33\%, 44\%, and 25\% respectively~\cite{aws-price-drop}.
% Even in major cloud provider such as AWS, in 2025 alone, the A100, H100, and H200 on-demand price have dropped by 33\%, 44\%, and 25\% respectively~\cite{aws-price-drop}.

\subsection{Software Trend}
\label{sec:software-trend}

Another barrier in building GPU DBMSes is the engineering cost and expertise required to build the software stack from the ground up. 
% This is particularly challenging as CPUs and GPUs have fundamental hardware differences. 
Fortunately, the barrier has been significantly lowered with the following two software trends. 

\spara{Composable Data Systems} Modern open-source data systems are embracing composability~\cite{composable} (e.g. Velox~\cite{velox}), where different system components can be independently developed and interoperate in a data system. A new database can reuse existing components such as the language frontend, query optimizer, or execution engine instead of reinventing the wheel. 
A crucial component in these systems is Substrait~\cite{substrait}, a unified IR for physical and logical representations of query plans. 
The trend of composable data systems significantly lowers the barrier to build a GPU database---only the execution engine needs to be developed from the ground up tailored for GPU but other system components could be reused. 

\spara{Maturing GPU Libraries}
Open-source GPU libraries optimized for data processing have emerged and are rapidly reaching maturity in recent years. 
For example, libcudf~\cite{cudf} provides GPU-efficient implementation for relational operators such as joins, aggregations, and sorting. RMM~\cite{rmm} complements this by offering a high-performance GPU memory allocator, serving as a foundation for GPU-centric buffer managers. On the distributed side, the NCCL~\cite{nccl} provides high-speed communication between GPU devices over various interconnect, such as PCIe, NVLink, and Ethernet/RDMA. 

%% file: sirius.tex
\section{Sirius: A GPU-Native SQL Engine}
\label{sec:sirius}

Motivated by the hardware and software trends, we have designed Sirius, a GPU-native SQL engine that provides drop-in acceleration across a variety of existing data systems. In this section, we will discuss the design overview and system architecture of Sirius.

\begin{figure}[t]
    \includegraphics[width=0.92\columnwidth]{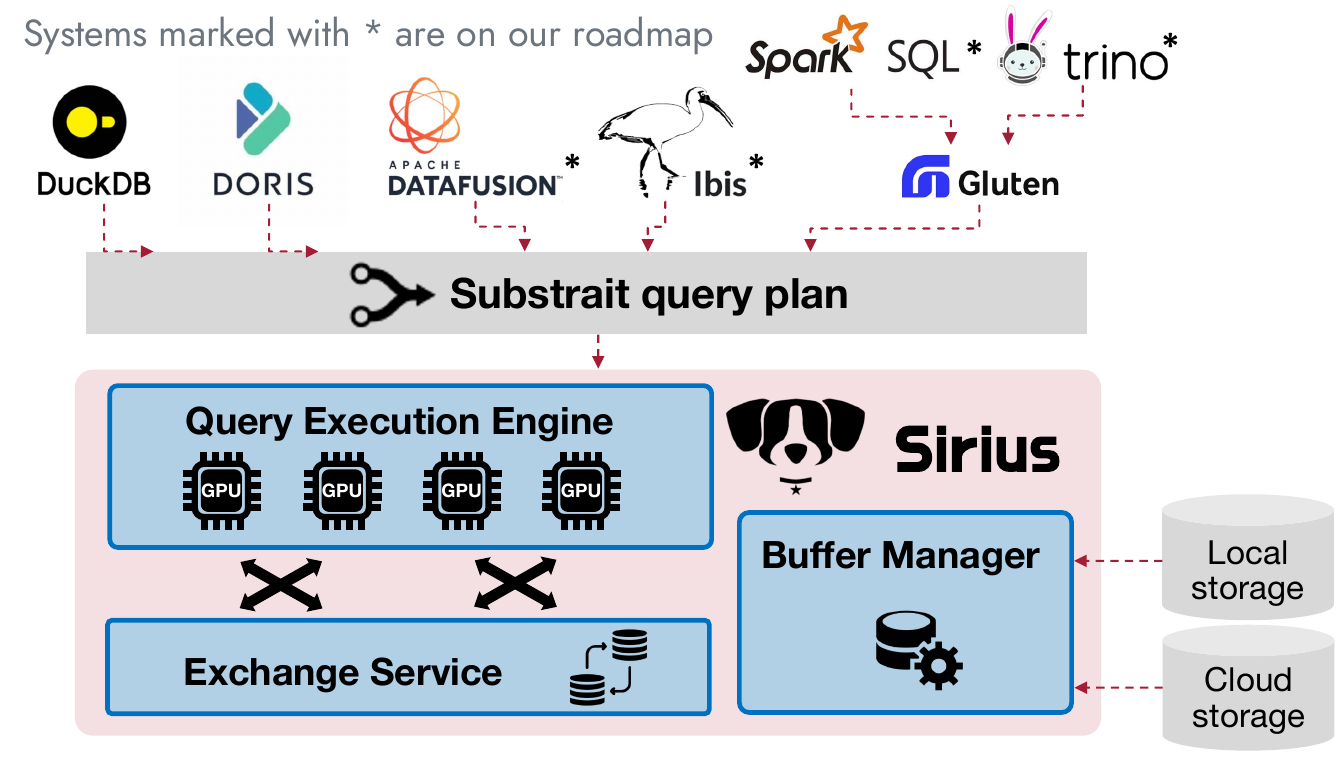}
    \vspace{-.15in}
    \caption{Sirius Architecture}
    \vspace{-.15in}
    \label{fig:sirius-architecture}
\end{figure}

\subsection{Design Overview}
\label{sec:design-overview}

%Taking into account the hardware and software trend discussed in Section~\ref{sec:motivation}, we 
Sirius was designed with the following two key design principles:

\spara{GPU-Native Execution}
Recent hardware trends show that the overhead of data movement between GPUs and other system components—especially CPUs—is diminishing. As the GPU memory barrier is breaking down, GPU-only query execution is becoming increasingly competitive with fine-tuned CPU–GPU co-execution.

% Taking this into account, we design Sirius as a \textbf{\textit{GPU-native engine}}: it treats GPUs as the primary execution engine, aiming to run the entire query plan—from scan to result—on the GPU. Unlike hybrid engines~\cite{mordred, maximus, lancelot, hetexchange, hetcache}, Sirius avoids splitting execution between CPU and GPU. Instead, it builds its execution model, buffer management, exchange service, and operator implementations entirely around full GPU residency. The CPU is used only as a fallback path when certain features are not supported on GPUs.

% Additionally, by designing Sirius as a GPU-native engine model, we avoid the complexity of coordinating heterogeneous execution and generating hybrid query plans, thereby improving portability.

Taking this into account, we design Sirius as a \textbf{\textit{GPU-native engine}}: it treats GPUs as the primary execution engine, aiming to run the entire query plan—from scan to result—on the GPU. Sirius differs from systems that retrofit GPU acceleration onto traditional CPU-optimized engines while maintaining backward compatibility~\cite{sparkrapids, velox-cudf}, or hybrid systems that split execution between CPU and GPU~\cite{mordred, maximus, lancelot, hetexchange}. Instead, it builds its execution model, buffer management, and exchange mechanisms entirely around full GPU residency. The CPU serves only as a fallback path when certain features are not supported on GPUs. 
This architecture can reduce the design complexity compared to a retrofitted or heterogeneous design, thereby improving portability and performance.

We believe GPU-native design is a promising direction. History shows that when a major infrastructure shift occurs—such as the rise of the cloud—systems built natively for the new paradigm (e.g., Snowflake) ultimately surpass retrofits of prior-generation architectures. Similarly, as GPUs emerge as a central computational platform for data analytics, GPU-native systems like Sirius have the potential to outperform hybrid or retrofitted approaches.

% Additionally, by designing Sirius as a GPU-native engine model, we avoid the complexity of coordinating heterogeneous execution and generating hybrid query plans, thereby improving portability.

\spara{Drop-In Acceleration}
Modern software stacks are gradually moving toward composable architectures~\cite{composable}, where components such as query optimizers, storage format, and query execution engine are decoupled. 
Sirius embraces this modular design to significantly reduce the development cost and the barrier of adoption.

By consuming a universal query plan format (e.g., Substrait) and connects it with existing GPU-accelerated libraries (e.g., libcudf), Sirius can serve as a \textit{\textbf{drop-in accelerator}} across a variety of data systems with minimal or no modification to the host systems.
When using Sirius, the users can benefit from GPU acceleration without having to modify their user interface or migrate to a different DBMS.
% , thereby lowering the barrier of adoption.

\subsection{System Architecture}
\label{sec:system-architecture}

In this Section, we will discuss the system architecture of Sirius as shown in Figure~\ref{fig:sirius-architecture}:
% provides an overview of the architecture, which consists of four main layers:

\subsubsection{Host Database Layer} \

This layer includes the query parser and optimizer of external data systems that Sirius accelerates. These host systems generate either query plans in a standardized format such as Substrait which Sirius can consume. Currently, Sirius supports DuckDB and Apache Doris as external hosts, and more will be supported in the future.

%\bwy{yifei should check this}
For distributed query execution, Sirius leverages the host database’s coordinator to manage the control plane. This includes identifying active nodes via heartbeat, scheduling plan fragments across nodes, determining partitioning schemes, and maintaining global metadata. 
For data exchange, however, it is handled by Sirius built-in exchange service layer explained in Section~\ref{sec:exchange-service}.
%\yxy{also briefly talk about how we handle data exchange.}
% Sirius bypasses the host databases distributed query execution through local Sirius execution engines (Section~\ref{sec:query-execution-engine}), and a GPU-native exchange service layer built on top of NCCL communication primitives (Section~\ref{sec:exchange-service}).

DuckDB supports Substrait and has an extension framework, allowing us to integrate Sirius with \textit{zero modification} to DuckDB's codebase.
For Apache Doris, 
% since it cannot export Substrait plans,
since does not have extension framework and cannot export Substrait plans. 
% To accelerate Doris, 
we introduce minimal modifications to its codebase to convert its internal query plans into the Substrait format.

% DuckDB supports Substrait and has an extension framework, allowing us to implement Sirius with \textit{zero} modification to DuckDB's codebase.
% In contrast, Doris lacks both an extension framework and support for Substrait, therefore, we introduce minimal modifications to its codebase to translate its query plans into Substrait.

\subsubsection{Query Execution Engine} \
\label{sec:query-execution-engine}

The query execution engine in Sirius executes the Substrait plan generated by the optimizer. Sirius adopts a \textbf{pipeline execution model}, in which the query plan is divided into pipelines. Each task—at the granularity of a pipeline—is enqueued into a global task queue. Idle CPU threads pull tasks from this queue and execute the pipelines by invoking each operator. This model is widely-used in modern data systems such as DuckDB, Hyper, and Velox.
%\kck{My understanding is that morsel-driven parallelism has more to do with multi-threaded execution: different threads pull morsel-sized tasks off of a shared task queue. DuckDB does this, but I don't think we do, at least not on GPU, since there is a single stream of execution.}

% Within each pipeline, Sirius adopts a \textbf{push-based execution model}~\cite{duckdb}. The query executor acts as a global coordinator, maintaining the state of query execution (e.g., operator's input and output) for each pipeline. Rather than having operators \textit{pull} input data from their predecessors, the executor \textit{pushes} data operators instead. This design keeps the operators stateless, simplifying their implementation complexity. The push-based execution model is also adopted by modern systems such as DuckDB.

Within each pipeline, Sirius adopts the \textbf{push-based execution model}. The query executor acts as a coordinator, maintaining the state of query execution (e.g., operator's input and output). Instead of having operators \textit{pull} data from their predecessors --as done in systems like Velox-- the executor \textit{pushes} data into operators. This design keeps the operators stateless, simplifying their implementation. This model is adopted by modern systems such as DuckDB.

Most physical operators in Sirius are implemented using the \texttt{libcudf} library, with a few exceptions for specialized functions such as predicate pushdown, and materialization. Thanks to its modular design, Sirius allows developers to easily switch the operator implementation between \texttt{libcudf} and custom CUDA kernels.
Sirius also includes a graceful fallback mechanism to the host database systems in the case of an error or missing features in Sirius. 
% Future extensions such as batch/pipeline execution and multi-GPU support will be discussed in Section~\ref{sec:future-extension}.

% Intermediate query results can be large and memory-intensive, placing pressure on the limited capacity of GPU memory. To alleviate this, Sirius adopts a \textbf{late materialization} strategy. Instead of immediately producing full intermediate results, the outputs of filters and joins are represented as row IDs and are only materialized when used by downstream operators. This approach reduces memory usage and can significantly improve performance, particularly for selective queries.

\subsubsection{Buffer Manager} \
\label{sec:buffer-manager}

The buffer manager is responsible for managing memory 
in Sirius. It divides the memory into two separate regions:

\spara{Data Caching}
This region is used to store cached data in Sirius either in device memory or pinned host memory. To minimize the overhead of dynamic memory allocation during query execution, the caching region is pre-allocated in advance. 
% While device memory provides faster access, pinned host memory typically offers greater capacity at the cost of slower throughput and latency.

\spara{Data Processing} 
This region is in device memory which stores intermediate results during query execution (e.g., hash tables, intermediate results, etc.). Sirius uses the RMM~\cite{rmm} (RAPIDS Memory Manager) pool allocator to manage this region efficiently. 

Moreover, the buffer manager is responsible for converting between different columnar formats used throughout the system: the internal Sirius columnar format, the \texttt{libcudf} columnar format, and the columnar format used by the host database. Both Sirius and libcudf derive their columnar format from Apache Arrow, which allows for zero-copy conversion via pointer passing.
The only exception is when converting row indices between Sirius and \texttt{libcudf}, as they use different types; Sirius uses \texttt{uint64\_t} whereas \texttt{libcudf} uses \texttt{int32\_t}.
Conversion between Sirius and the host database format also requires deep copies, but this occurs only during the cold run (initial data load) and when returning results to the user.

Currently, Sirius relies on the host database to read data from disk. Once data is read, the buffer manager automatically caches it into the pre-allocated caching region for future reuse. More discussion on disk support and spilling will be discussed in Section~\ref{sec:future-extension}.

\subsubsection{Exchange Service Layer} \
\label{sec:exchange-service}

The exchange service layer orchestrates distributed query execution across multiple nodes. In single-node deployments, this layer can be bypassed entirely. However, in multi-node settings, it plays a critical role in coordinating data exchange between nodes.

In Sirius, exchange is modeled as dedicated physical operators. Sirius supports common exchange patterns—\texttt{broadcast}, \texttt{shuffle}, \texttt{merge}, and \texttt{multi-cast}—all implemented using NCCL primitives. 

In distributed execution, the query plan is divided into multiple fragments, each of which is executed locally on a participating node. Data exchange occurs between fragments—after one fragment completes, its output is transmitted across the compute nodes and consumed as input by the subsequent fragments.
Sirius internally maintains a runtime registry of exchanged intermediate data as temporary tables. These temporary tables are automatically deregistered once the corresponding fragments finish execution.

% \noindent \bwy{Yifei feel free to add more here}

\subsection{Query Lifecycle in Sirius}
\label{sec:lifecycle}
In this section, we will describe the end-to-end query lifecycle when Sirius is used as a drop-in replacement for DuckDB (in single-node settings) and Apache Doris (in distributed environments).

% \spara{Distributed:} Figure~\ref{fig:sirius-dist} compares the query workflow between the Apache Doris (Figure ~\ref{fig:sirius-dist}(a)) and GPU-accelerated Doris (Figure ~\ref{fig:sirius-dist}(b)). Initially, the user input SQL query via Doris frontend. In Figure~\ref{fig:sirius-dist}(a), the Doris coordinator (i.e., frontend) generates the optimized distributed query plan and dispatch plan fragments to all compute nodes. The compute nodes then execute plan fragments locally, and exchange data through Doris data exchange service implemented on top of bRPC~\cite{brpc}.

% Figure~\ref{fig:sirius-dist}(b) demonstrates distributed query execution of GPU-accelerated Doris. 
% Sirius keeps the Doris coordinator unmodified to produce the exactly same distributed query plan, and dispatch plan fragments to all Doris backends. 
% After this, instead of directly executing the received plan fragments, each Doris backend converts internal Doris plan fragments into Substrait format, and forwards the substrait plans to Sirius execution engine (Section~\ref{sec:query-execution-engine}). On top of the Sirius execution engine at each node, Sirius deploys a GPU-native exchange service built on top of NCCL communication primitives, to perform exchange over intermediate output of individual plan fragments locally at each node. When the query finishes, the Sirius execution engines sends the query result to Doris backends.

\spara{Single-Node} Users begin by issuing SQL queries through the DuckDB CLI or Python API. DuckDB then parses and optimizes the query, producing a DuckDB-specific logical plan. In the vanilla DuckDB workflow, DuckDB-native execution engine will execute this plan and return the result to the user. When accelerated by Sirius, however, DuckDB delegates the execution to Sirius by passing the query plan to its GPU-accelerated execution engine (Section~\ref{sec:query-execution-engine})
% . Once completes, the buffer manager converts the result back to DuckDB’s internal format 
and returns the result to the user.

% \noindent \bwy{yifei, please check if this makes sense to you}

% \begin{figure}
% \subfloat[Vanilla Doris]{
%   \centering
%   \includegraphics[height=3.9cm]{figures/vanilla-doris.png}
% }
% \hfill
% \subfloat[Sirius-Accelerated Doris]{
%   \centering
%   \includegraphics[height=4.2cm]{figures/sirius-doris.png}
% }
% \vspace{-.12in}
% \caption{Distributed Query Lifecycle in Doris vs Sirius.}
% \vspace{-.15in}
% \label{fig:sirius-dist}
% \end{figure}

\begin{figure}[t]
    % \centering
    % First row
    \begin{subfigure}[b]{0.4\columnwidth}
        \centering
        \includegraphics[width=\columnwidth]{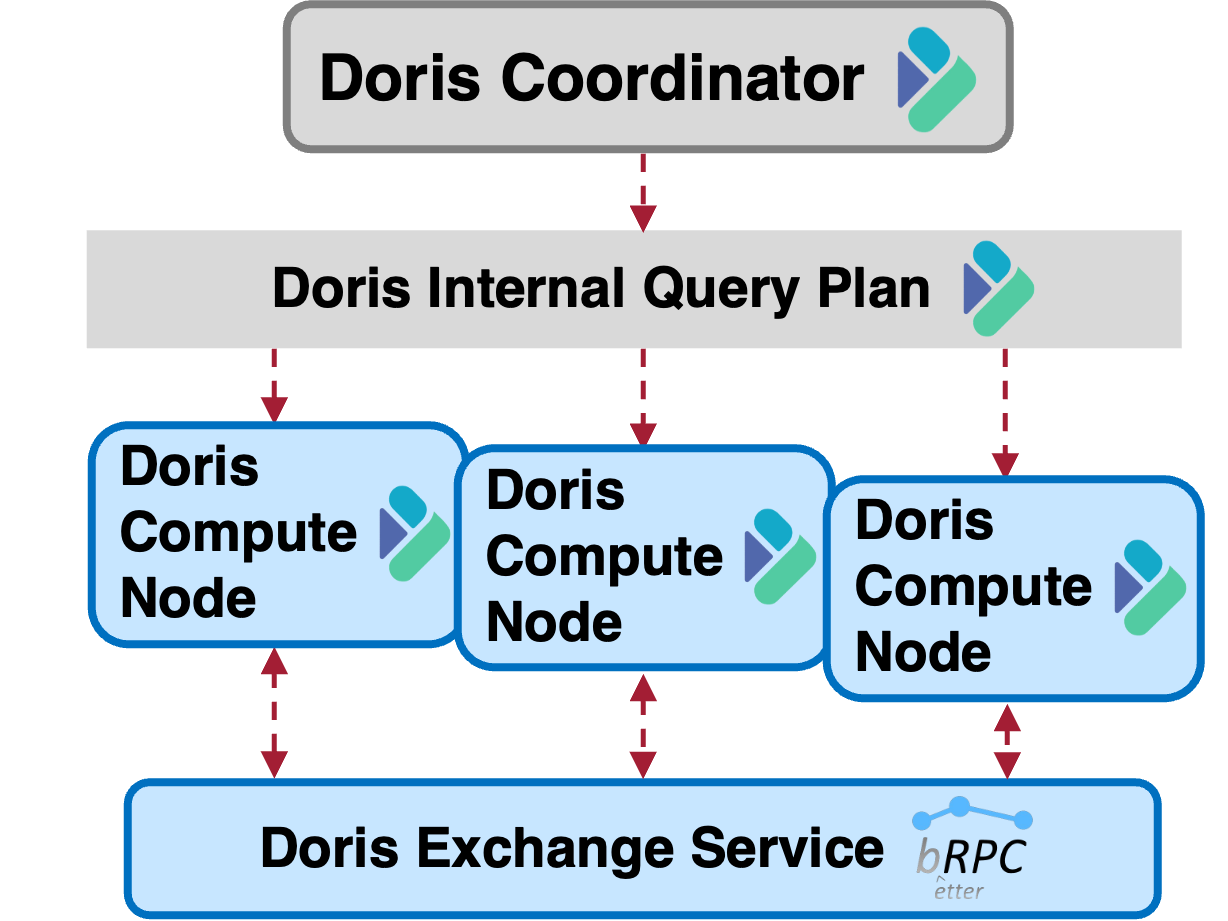}
        % \vspace{-.2in}
        \caption{Vanilla Doris}
    \end{subfigure}
    % \hfill
    \begin{subfigure}[b]{0.5\columnwidth}
        \centering
        \includegraphics[width=\columnwidth]{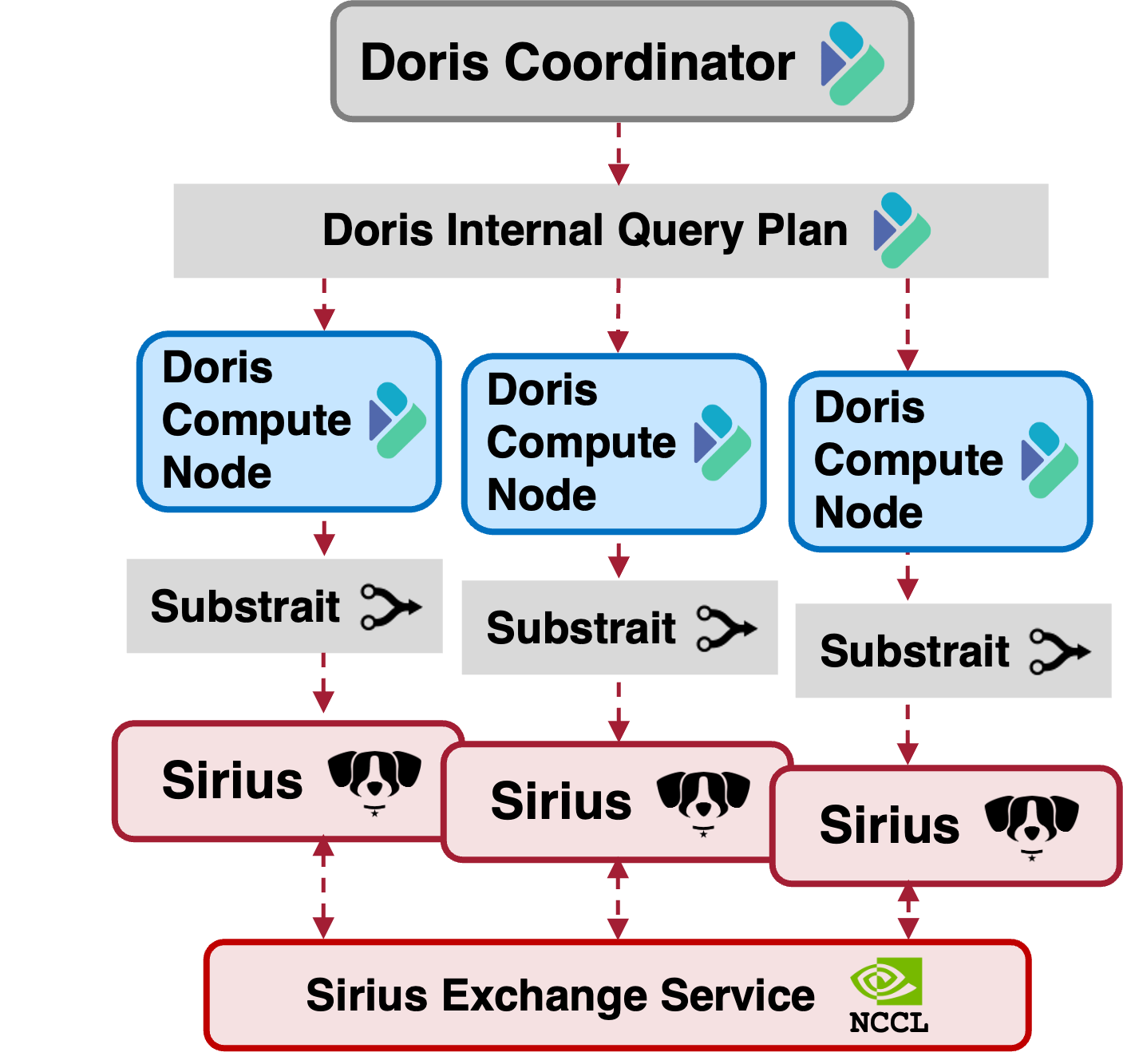}
        % \vspace{-.2in}
        \caption{Sirius-Accelerated Doris}
    \end{subfigure}
    \vspace{-.13in}
    \caption{Distributed Query Lifecycle in Doris vs Sirius.}
    \vspace{-.22in}
    \label{fig:sirius-dist}
\end{figure}

\spara{Distributed} Figure~\ref{fig:sirius-dist} illustrates the query lifecycle for vanilla Apache Doris (Figure~\ref{fig:sirius-dist}(a)) and GPU-accelerated Doris powered by Sirius (Figure~\ref{fig:sirius-dist}(b)). 
In both cases, users submit SQL queries via the Doris SQL interface. The Doris coordinator generates a distributed query plan and dispatches plan fragments to Doris compute nodes. 
In standard Doris workflow, the compute nodes will execute the fragments locally, exchange intermediate data via Doris' native exchange service, and return the results to the coordinator.

When accelerated by Sirius, instead of executing the plan fragments directly,
% Doris compute nodes 
% no longer execute the plan fragments directly. 
% Instead, 
each compute node translates its assigned fragment into Substrait and forwards it to the local Sirius execution engine (Section~\ref{sec:query-execution-engine}).
To exchange intermediate results across nodes, Sirius uses its exchange service layer described in Section~\ref{sec:exchange-service}. 
Once execution completes, Sirius returns the results to Doris compute node, which then forward the results to the coordinator.

\begin{figure*}[ht!]
    \centering
    \includegraphics[width=0.95\linewidth]{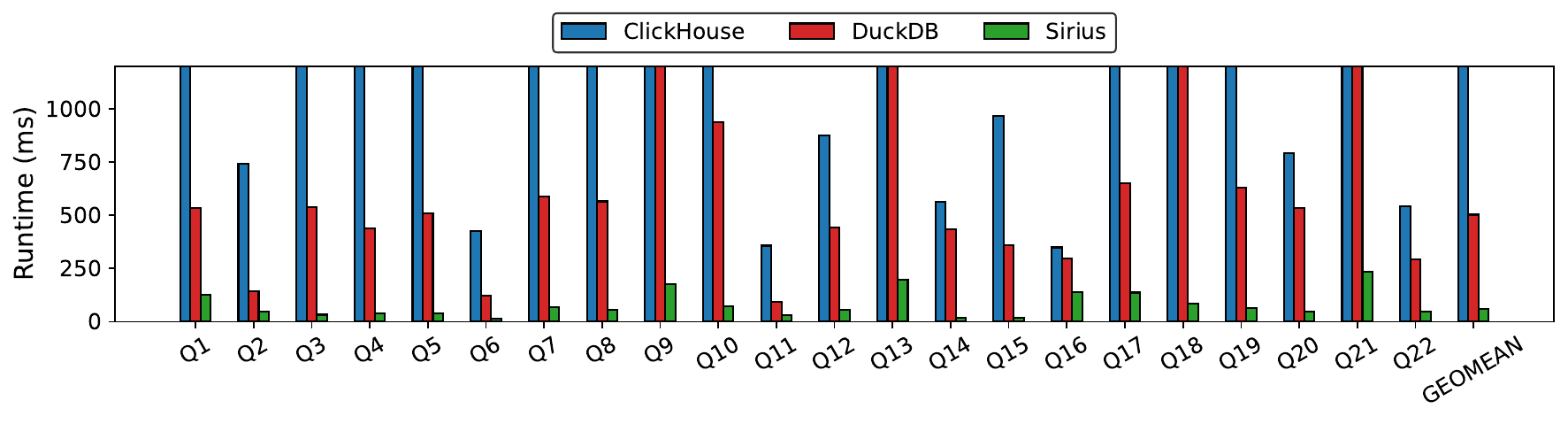}
    \vspace{-.3in}
    \caption{TPC-H End-to-end Query Performance on a Single Node.}
    \vspace{-.16in}
    \label{fig:endtoend}
\end{figure*}

\begin{figure}[ht!]
    \includegraphics[width=\columnwidth]{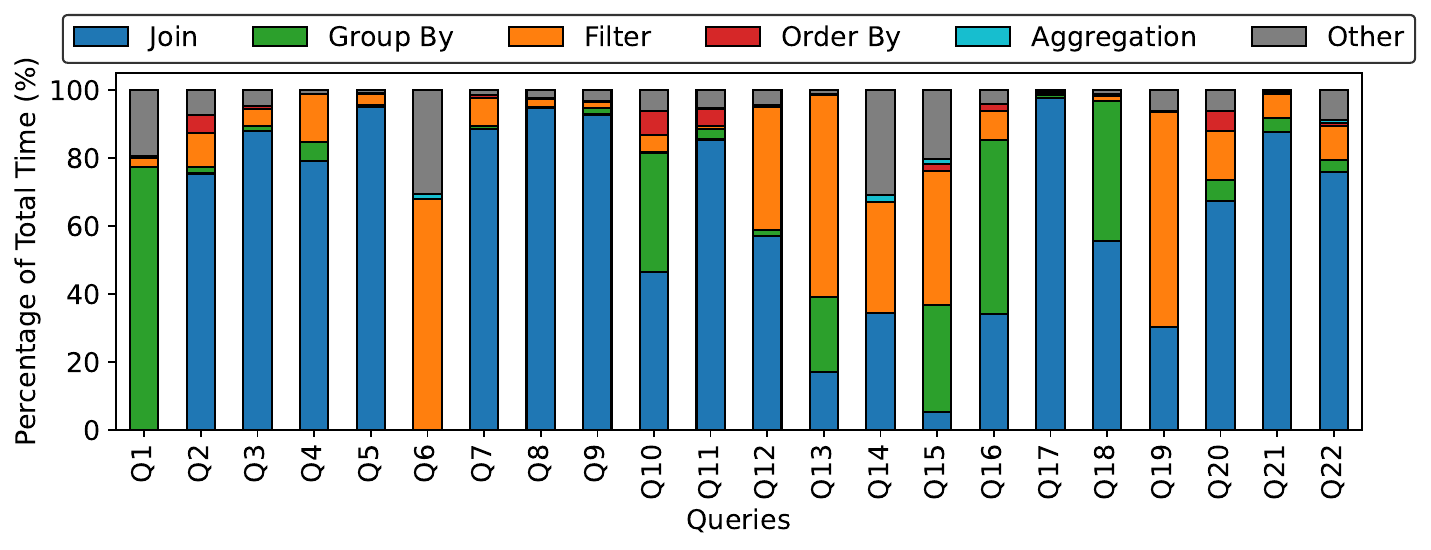}
    \vspace{-.33in}
    \caption{Performance Breakdown in Sirius}
    \vspace{-.15in}
    \label{fig:breakdown}
\end{figure}

\subsection{Future Extensions}
\label{sec:future-extension}

% We built Sirius to demonstrate that a GPU-native query engine can be composable, performant and pluggable into existing data systems. 
While our initial prototype targets the in-memory setting with limited feature sets, 
we plan to extend Sirius into a more complete and production-ready system. 

\spara{Out-of-Core Execution}
Currently, Sirius operates under the assumption that the input data fits in the caching region and the intermediate results fit in the processing region. To support larger-than-memory workloads, we plan to extend the buffer manager to support spilling to pinned memory and disk. We will also enhance the execution engine to support batch execution by partitioning input data into batches and pipelining them into the GPU~\cite{vortex, hetexchange}.

\spara{Full Distributed and Multi-GPU Support}
The current distributed mode offers limited SQL coverage compared to its single-node counterpart. For instance, it does not support functions such as \texttt{avg}. Going forward, we plan to expand the SQL coverage for the distributed execution~\cite{theseus} and introduce other key features such as fault tolerance. Additionally, we will extend Sirius to support multiple GPUs per node~\cite{lancelot}, enabling it to handle larger workload.
% and fully leverage modern GPU servers.

\spara{Expanding SQL Coverage}
Sirius already supports most of the basic SQL operators and data types. We plan to broaden the coverage by adding support for more complex data types, such as LIST, and nested types. Additionally, we aim to implement advanced SQL operators, including ASOF joins, vector search, and UDFs~\cite{gpu-udaf}.

\spara{Expanding Host Database Coverage}
Sirius currently serves as a drop-in accelerator for both DuckDB and Apache Doris. Going forward, we plan to support other substrait-compatible systems, such as DataFusion, Ibis, and Gluten~\cite{gluten}.

\spara{Performance Optimizations}
Sirius is still in its early stages, and we see significant opportunities for performance improvement. Future optimizations include predicate transfer~\cite{predicate-transfer, dist-pt}, efficient distributed shuffling, operator-specific enhancements~\cite{gpu-join-eth}, and lightweight compression techniques~\cite{gpu-compression, fastlane-gpu} to mitigate GPU memory capacity limitations—particularly during batch/pipeline execution~\cite{theseus, golap}.
Additionally, we aim to optimize I/O paths for reading data from disk and across the network using GPUDirect~\cite{gpudirect}.

% As an open-source project, Sirius welcomes contributions and collaboration from the database community. We invite researchers and practitioners to build on top of Sirius, with the shared mission of driving the next era of data analytics.

%% file: evaluation.tex
\begin{figure*}[ht!]
    \centering
    \includegraphics[width=0.9\linewidth]{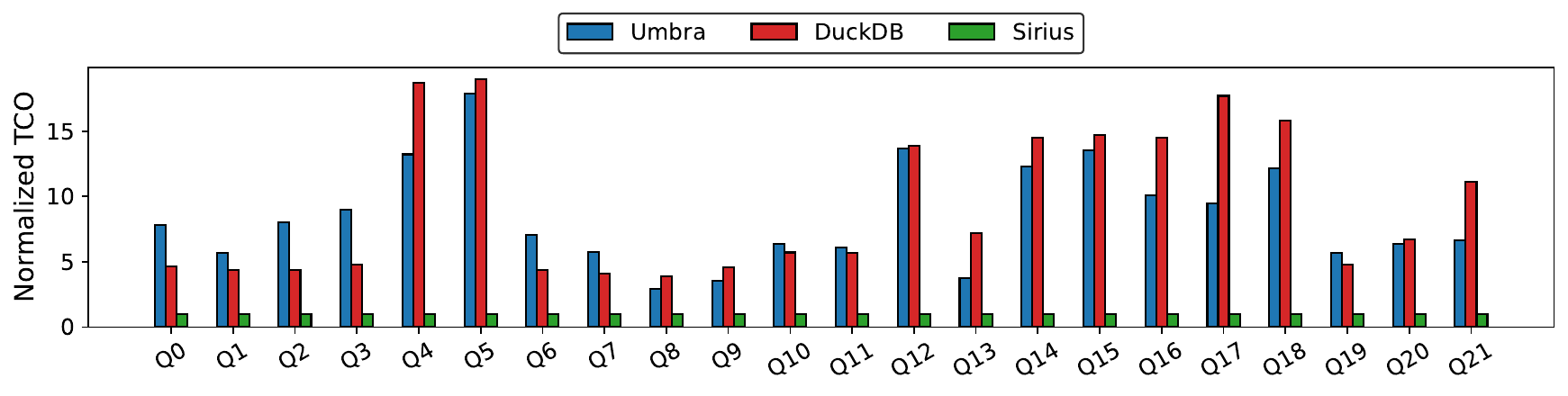}
    \includegraphics[width=0.9\linewidth]{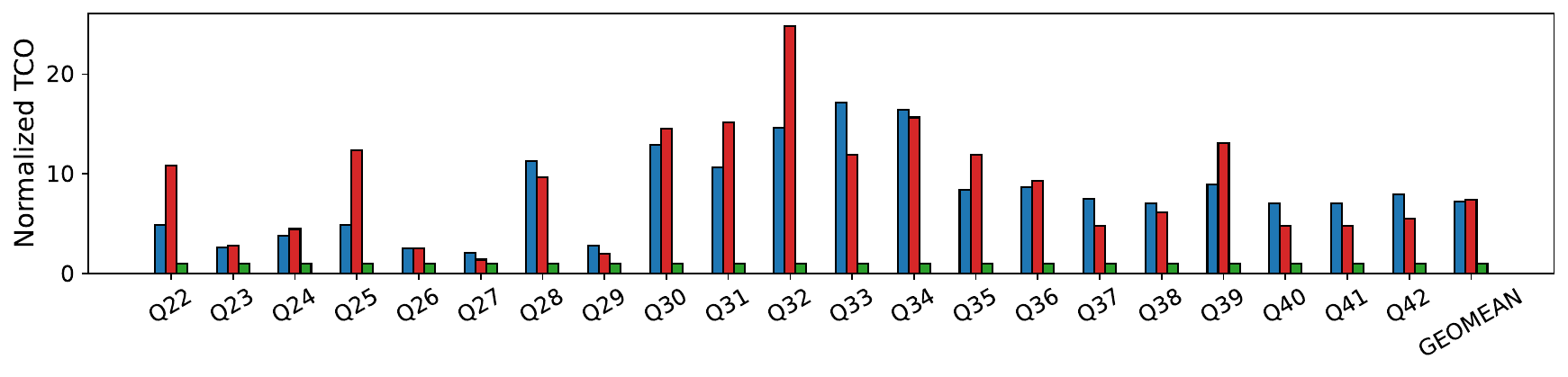}
    \vspace{-.3in}
    \caption{Normalized Total Cost of Ownership (TCO) Relative to Sirius.}
    \vspace{-.16in}
    \label{fig:clickbench}
\end{figure*}

\section{Evaluation}
\label{sec:evaluation}
In this section, we will evaluate the performance of Sirius as a drop-in accelerator for DuckDB (single-node) and Doris (distributed).

\subsection{Experiment Setup}
\label{sec:experiment-setup}

\spara{Hardware configuration} We ran Sirius with two different setups:

\setlist[itemize]{align=parleft,left=0pt..1em}
\begin{itemize}
    \item \textbf{NVIDIA GH200:} This machine features NVIDIA Grace CPU and Hopper GPU connected through NVLink C2C with 900~GB/s bidirectional bandwidth. The Hopper GPU has a 92~GB of HBM3 memory with a read/write bandwidth of 3~TB/s. The Grace CPU has 72 Neoverse Armv9 cores with 480~GB of LPDDR5X memory.
    \item \textbf{4 $\times$ NVIDIA A100:} This cluster consists of 4 identical nodes, where each node has one NVIDIA A100 GPU and 64-cores Intel Xeon Gold 6526Y CPUs. The A100 GPU has 40~GB of HBM3 memory with a read/write bandwidth of 1550~GB/s. The GPU and CPU are connected via PCIe4 with 25.6~GB/s bidirectional bandwidth. Every node support InfiniBand 4x NDR (Next-Generation Rate), providing network bandwidth of up to 400~Gbps.
\end{itemize}

\spara{Benchmark} We use the \textit{TPC-H} benchmark and \textit{ClickBench}~\cite{clickbench} in our experiment, which has been widely used both in academia and industry. 
% We will run the full-set of TPC-H queries in the single-node setting %  Section~\ref{sec:single-node} 
% and a subset of TPC-H queries in the distributed setting (Q1, Q3, and Q6). 
%Section~\ref{sec:distributed}.

\spara{Measurement} 
We dedicate $50\%$ of each GPU memory for data caching, and the other half for data processing. \textit{The numbers reported are the hot runs when the data is already \textbf{cached in the GPU memory}}. Throughout the experiment, we will evaluate the performance of the following query engines:

\begin{itemize}
    \item \textbf{DuckDB: } DuckDB~\cite{duckdb} is a single node open-source embedded CPU-based analytical database.
    \item \textbf{ClickHouse: } ClickHouse~\cite{clickhouse} is a distributed open-source CPU-based online analytical database.
    \item \textbf{Umbra: } Umbra~\cite{umbra} is a single-node RDBMS designed to process data at line-speed for both in-memory or storage resident data.
    \item \textbf{Apache Doris: } Doris~\cite{doris} is an open source high-performance and real-time distributed CPU-based data warehouse.
    \item \textbf{Sirius: } Our GPU-native SQL engine that serves as drop-in accelerator for DuckDB (single node) and Apache Doris (distributed).
\end{itemize}

%In Section~\ref{sec:single-node}, we compare Sirius against DuckDB, ClickHouse, and HeavyDB; In Section~\ref{sec:distributed}, we compare Sirius agains Doris and ClickHouse.

\subsection{TPC-H Single-Node Performance}
\label{sec:single-node}

In this Section, we evaluate Sirius performance running on a single node as a drop-in accelerator for DuckDB. We ran 22 TPC-H queries with a scale factor of 100 and compare the performance against DuckDB and ClickHouse. 
To ensure cost-normality, we ran DuckDB and ClickHouse on a CPU instance (c8i.8xlarge@AWS) that has the same hourly rental cost (\$1.5/h) as the GPU instance (GH200@LambdaLabs) used to run Sirius. %and HeavyDB.

\spara{End-to-end Performance Comparison}
Figure~\ref{fig:endtoend} shows the end-to-end performance of various query engines on the TPC-H benchmark. Compared to DuckDB, Sirius achieves a 8.3$\times$ speedup with the same hardware rental cost. 
Sirius leverages DuckDB’s optimized logical plans but replaces its backend with GPUs, demonstrating the significant performance advantage of GPU acceleration.

Since ClickHouse does not support correlated subqueries~\cite{clickhouse}, we rewrite queries containing subquery correlation for compatibility. Compared to ClickHouse, Sirius is 38$\times$ faster with the same hardware rental cost. Q9 does not finish in ClickHouse and Q21 is not supported. We observed that ClickHouse is not optimized for join-heavy workloads, leading to suboptimal performance in many TPC-H queries (e.g., Q2, Q5, Q10, etc).

% Compared to HeavyDB, Sirius is a 1.5$\times$ faster. HeavyDB is a standalone GPU database system built from the ground up and employs JIT compilation to generate efficient GPU kernels. In contrast, Sirius is a GPU-accelerator for existing database systems and is built by leveraging composable data systems stack. Despite its modular design, Sirius is still able to outperform HeavyDB, demonstrating the effectiveness of modern, reusable data systems components.

\spara{Performance Breakdown}
Figure~\ref{fig:breakdown} presents a performance breakdown across different TPC-H queries in Sirius. The results indicate that join operations dominate query execution time in most cases, particularly in join-heavy queries (Q2–Q5, Q7–Q8, Q20–Q22).

Group-by also account for a substantial portion in several queries, notably for Q1, Q10, Q16, and Q18. This overhead is especially visible in queries involving group-by on string keys (Q10, Q18) or with a small number of distinct groups (Q1). For string keys, libcudf defaults to a sort-based group-by strategy, which is less performant than hash-based group-by. For group-by operations with a small number of groups, GPU suffers from memory contention.

Filter operation contributes to a majority of the query execution time in filter-heavy queries such as Q6 and Q19. They also contribute significantly in Q13, where a complex string-matching expression with very low selectivity is used. In contrast, aggregation and order-by operations do not dominate the end-to-end execution time for any of the TPC-H queries. This is because the input size for these operations is typically very small.%smaller than that of the input tables, as TPC-H queries tend to be highly selective.

\subsection{TPC-H Distributed Performance}
\label{sec:distributed}

%In this section, we evaluate Sirius performance running on a distributed environment (a cluster of as a drop-in accelerator for Apache Doris.
In this section, we ran a subset of TPC-H queries with a scale factor of 100 (i.e., Q1, Q3, and Q6) that are currently supported by the distributed mode of Sirius and compare the performance of Sirius against Doris and ClickHouse using a cluster of 4 $\times$ A100 GPUs. 

\vspace{-.1in}
\begin{table}[ht]
  \caption{TPC-H End-to-End Query Performance in the Distributed Setting.}
  \vspace{-.1in}
  \label{tab:endtoend-dist}
  \small
  \begin{tblr}{
    colsep=4pt,
    colspec = {c|ccc|ccc}
  }
    \toprule
    \SetCell[r=2]{c} & \SetCell[r=2]{c} Doris & \SetCell[r=2]{c} ClickHouse & \SetCell[r=2]{c} Sirius & \SetCell[c=3]{c} Breakdown in Sirius\\
    \cmidrule{1-7}
    & & & & Compute & Exchange & Other\\
    \cmidrule{1-7}\morecmidrules\cmidrule{1-7}
    Q1 & 1193 & 393 & \textbf{97} & 33 & 3 & 61\\
    \cmidrule{1-7}
    Q3 & 838 & 12785 & \textbf{341} & 43 & 233 & 75\\
    \cmidrule{1-7}
    Q6 & 199 & 294 & \textbf{84} & 36 & 1 & 47\\
    \bottomrule
  \end{tblr}
\end{table}

\spara{End-to-end Performance Comparison} Table~\ref{tab:endtoend-dist} presents the performance comparison of different query engines in a distributed environment. Compared to Doris, Sirius achieves a speedup of 12.5$\times$, 2.5$\times$, and 2.4$\times$ in Q1, Q3, and Q6, respectively. Sirius leverages the exact same distributed query plan from Doris, but replaces its execution engine with GPUs, demonstrating the performance advantage of GPUs in distributed query execution.

Sirius outperforms ClickHouse by 4$\times$, 37.5$\times$, and 3.5$\times$ in Q1, Q3, and Q6, respectively. When joins are not involved (Q1, Q6), ClickHouse is able to outperform Doris. However, its performance degrades significantly when distributed joins are involved (i.e., Q3), resulting in substantial slowdown compared to Doris and Sirius.

\spara{Performance Breakdown} We further present a breakdown of Sirius' performance in Table~\ref{tab:endtoend-dist}, specifically, into the categories of local GPU computation, data exchange, and the rest portion during execution (e.g., query optimization, plan dispatching, etc.).

We noticed that GPU execution is not the primary performance bottleneck in the current implementation of distributed Sirius, indicating ample opportunities for further optimizations. For example, in Q1 and Q6, a significant portion of query execution time is spent in Doris' query optimizer and coordinator, which run on the CPU. This overhead does not scale with the data size and is expected to be less significant for larger dataset. In Q3, data exchange is the main bottleneck as the Doris' distributed query plan shuffles both the \texttt{orders} and \texttt{lineitem} tables. We have identified several opportunities to further improve the shuffle performance in Sirius but leave a deeper exploration to future work. 

%We observe that for light queries like Q1 and Q6, the distributed execution (i.e., compute time and exchange time) within Sirius is not the major bottleneck. For example, on Q1 only 37\% of the end-to-end runtime is spent on the distributed execution, which correspond to 3\% and 9\% of the end-to-end execution time of Doris and ClickHouse, respectively. This demonstrates the high efficiency of Sirius's distributed query execution.

%On Q3, and we observe that data exchange is the major bottleneck. Q3 involves distributed joins and the query plan from the host system (i.e., Doris) requires to shuffle both \texttt{orders} and \texttt{lineitem} tables. We already identified the potential optimization opportunities in Sirius's data shuffle service, which is left as future work.

\subsection{ClickBench Performance}
\label{sec:clickbench}

In this section, we evaluate Sirius as a drop-in accelerator for DuckDB and compare its performance against the top two systems on the ClickBench leaderboard (Umbra and DuckDB\footnote{as of December 5, 2025; results are subject to change}). We ran Sirius on a GH200 instance from Lambda Labs (\$1.5/hour). The competing systems ran on CPU instances—AWS c8g.metal-48xl (\$7.6/hour) and AWS c7a.metal-48xl (\$9.8/hour). 
Because the three systems use differently priced hardware, we report the normalized total cost of ownership (TCO) relative to Sirius.

Figure~\ref{fig:clickbench} shows the normalized TCO of Umbra and DuckDB relative to Sirius (fixed at 1).
Sirius achieves the lowest TCO on all queries, benefiting from highly efficient GPU execution. In particular, queries q4, q5, and q18 exhibit substantial performance gains on common operators such as filtering, projection, and aggregation.
Some queries, however, highlight optimization opportunities. For example, q23 is bottlenecked by the contains operation on string columns; q24 and q26 by top-N operators; and q27 by aggregations over very large inputs. We plan to improve these operators in future releases of Sirius. 

Overall, Sirius delivers at least a 7.4× improvement in TCO when running ClickBench as a drop-in accelerator for DuckDB. It achieves the lowest relative runtime on the ClickBench leaderboard—11\% faster than Umbra, the top performer—while running on significantly cheaper hardware (\$1.5/hour vs. \$9.8/hour), setting a new performance record for the benchmark.

%% file: conclusion.tex
\section{Conclusion}
\label{sec:conclusion}

This paper argues that the GPU era for data analytics has arrived and supports this claim by examining recent trends in hardware and software development. 
Motivated by these trends, we introduce Sirius, an open-source GPU-native SQL engine that provides drop-in acceleration across a variety of existing data systems. 
Our evaluation on TPC-H and ClickBench shows that Sirius delivers 7.4–8.3$\times$ better TCO than state-of-the-art systems in a single-node setting, and up to 12.5$\times$ speedup in distributed setting.
With a permissive Apache-2.0 license, Sirius welcomes contributions from researchers and practitioners in the database community with the shared mission of kickstarting the GPU eras for data analytics.

%% file: acknowledgment.tex
\section{Acknowledgment}
\label{sec:acknowledgment}
This work was supported (in part) by the Lambda Cloud Credits and NVIDIA Data Science Cloud Credits. We thank Ming Liu for providing hardware access for our distributed experiments. We also thank Felipe Aramburu, Amin Aramoon, Rodrigo Aramburu, Matthijs Brobbel, Johan Peltenburg, and Dhruv Vats for their tremendous contributions to Sirius.